\documentclass[sigconf]{acmart}

\usepackage[utf8]{inputenc} 
\usepackage[T1]{fontenc}    
\usepackage{hyperref}       
\usepackage{url}            
\usepackage{booktabs}       
\usepackage{amsfonts}       
\usepackage{nicefrac}       
\usepackage{microtype}      
\usepackage{float}
\usepackage{natbib}
\usepackage{soul}
\usepackage{graphics}
\usepackage{graphicx}
\usepackage{subfigure}

\usepackage{url}
\usepackage{balance}

\usepackage[ruled,vlined,linesnumbered]{algorithm2e}
\usepackage{enumerate}
\usepackage{paralist}
\usepackage{multirow,multicol,xspace}
\usepackage{amsthm,amsmath}

\newtheorem{dfn}{Definition}

\RequirePackage{algorithmic}

\usepackage{cleveref}
\crefname{equation}{Eq.}{Eqs.}
\crefname{table}{Table}{Tables}
\crefname{figure}{Figure}{Figures}
\crefname{section}{Section}{Sections}
\crefname{algorithm}{Algorithm}{Algorithms}

\AtBeginDocument{%
  \providecommand\BibTeX{{%
    \normalfont B\kern-0.5em{\scshape i\kern-0.25em b}\kern-0.8em\TeX}}}

\setcopyright{acmcopyright}
\copyrightyear{2022}
\acmYear{2022}
\acmDOI{}

\acmConference[DS4RRS '22]{Workshop on Data Science and Artificial Intelligence
for Responsible Recommendations (DS4RRS)}{August 14-18, 2022}{Washington DC, USA}
\acmBooktitle{Workshop on Data Science and Artificial Intelligence
for Responsible Recommendations (DS4RRS), August 14-18, 2022, Washington DC, USA}
\acmPrice{}
\acmISBN{}

\begin{document}

\title{On the Relationship \\ Between Counterfactual Explainer and Recommender}

\author{Gang Liu, Zhihan Zhang, Zheng Ning, Meng Jiang} 
\affiliation{%
  \institution{{Department of Computer Science and Engineering, University of Notre Dame \country{USA}}}
}
\email{{gliu7, zzhang23, zning, mjiang2}@nd.edu }

\newcommand{\accentt}{\textsc{ACCENT}\xspace}
\newcommand{\fia}{\textsc{FIA}\xspace}
\newcommand{\retrain}{\textsc{FT}\xspace}

\begin{abstract}
Recommender systems employ machine learning models to learn from historical data to predict the preferences of users. Deep neural network (DNN) models such as neural collaborative filtering (NCF) are increasingly popular. However, the tangibility and trustworthiness of the recommendations are questionable due to the complexity and lack of explainability of the models.  To enable explainability, recent techniques such as ACCENT and FIA are looking for counterfactual explanations that are specific historical actions of a user, the removal of which leads to a change to the recommendation result. In this work, we present a general framework for both DNN and non-DNN models so that the counterfactual explainers all belong to it with specific choices of components. This framework first estimates the influence of a certain historical action after its removal and then uses search algorithms to find the minimal set of such actions for the counterfactual explanation. With this framework, we are able to investigate the relationship between the explainers and recommenders. We empirically study two recommender models (NCF and Factorization Machine) and two datasets (MovieLens and Yelp). We analyze the relationship between the performance of the recommender and the quality of the explainer. We observe that with standard evaluation metrics, the explainers deliver worse performance when the recommendations are more accurate. This indicates that having good explanations to correct predictions is harder than having them to wrong predictions. The community needs more fine-grained evaluation metrics to measure the quality of counterfactual explanations to recommender systems.
\end{abstract}

\maketitle

\section{Introduction}
\label{sec:introduction}
A recommendation system (RecSys) acts as a information filtering system that aims at predicting ratings or preference a user might have towards an item~\cite{jiang2014scalable, wang2018multi}. One of the most popular methods for generating recommendations is collaborative filtering, which aims at predicting users' preferences against a set of items based on past user-item interactions. In the era of deep learning, Neural Collaborative Filtering (NCF) becomes one of the most popular models used in RecSys. In NCF, the general idea is to first map users and items into a latent vector space via neural networks, then the model can compute the preference score for each user-item pair using the vectors. The model is trained on historical data of user-item interactions. However, though deep learning based methods usually deliver better performance compared with traditional methods, the model itself is a blackbox. As there is no intuitive meaning for each dimension of the latent vector, it is difficult for humans to understand the behaviors of the recommendation model.

\begin{figure*}[t]
\includegraphics[width=0.95\linewidth]{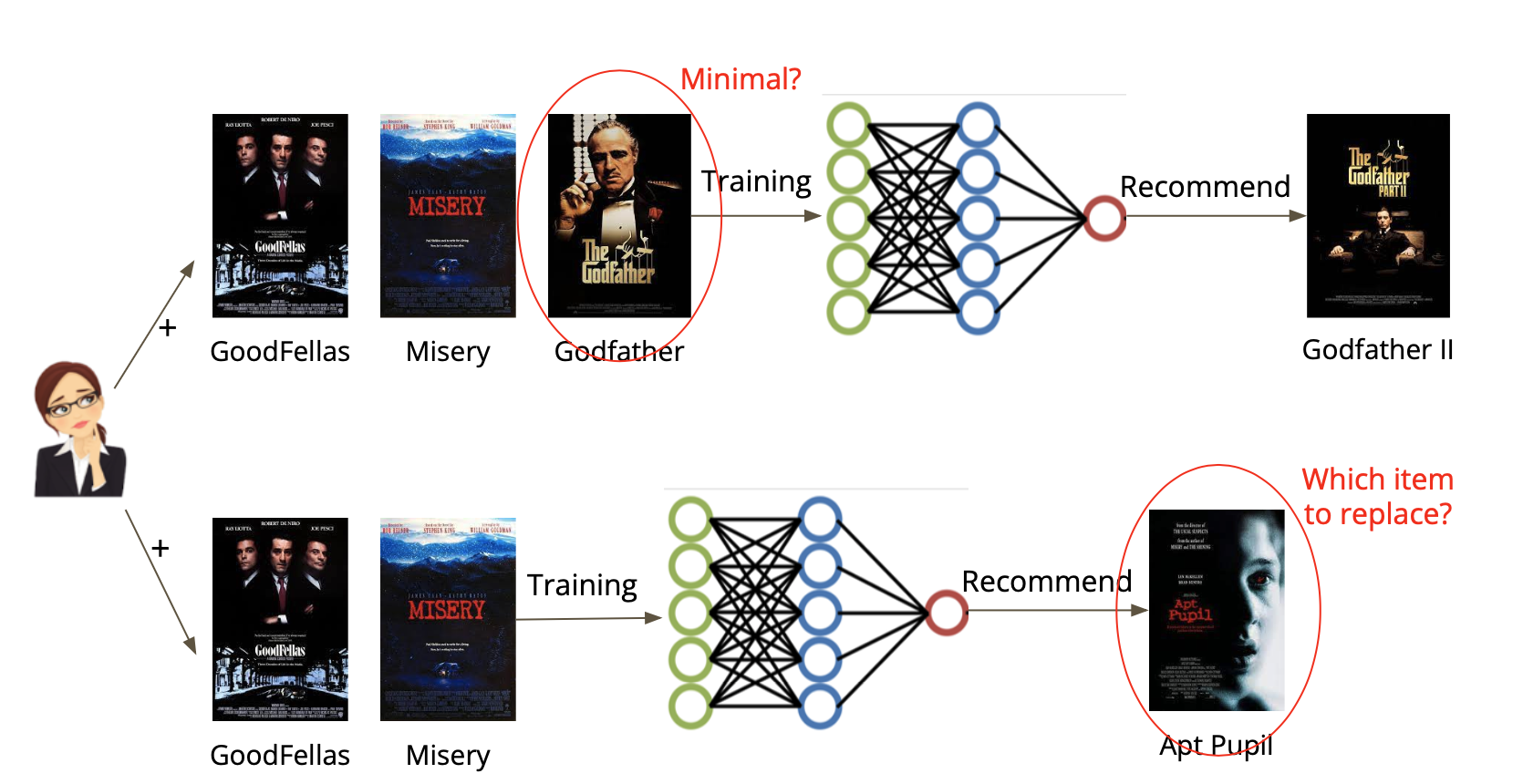}
\caption{Illustration of counterfactual explanation in neural recommender systems.}
\label{cf_explain}
\end{figure*}

It is known that explanations in RecSys should be actionable towards improving the underlying models and help enhance user experience~\cite{balog2020measuring,tran2021ACCENT}.
Many explanation models have been proposed to understand the intrinsic mechanism of RecSys~\cite{jiang2015social, lu2018coevolutionary,seo2017interpretable, abdollahi2016explainable,heckel2017scalable,yang2018towards}. The techniques in \cite{lu2018coevolutionary,seo2017interpretable} use external features, which are often not available in practice. The techniques in \cite{abdollahi2016explainable,heckel2017scalable,yang2018towards} search paths between users and items based on the similarity, which rises the concern of privacy and may not be actionable to users \cite{tran2021ACCENT}. 

Recently, the concept of counterfactual explanation \cite{tran2021ACCENT} is proposed to find scrutable and actionable explanations from the user's own actions for RecSys. The idea is basically to first predict the influence of the user’s actions on the recommendation results; and then to find a minimal set of actions, whose removal produces a different recommendation.
Figure \ref{cf_explain} shows an example. Suppose a user has watched \emph{Goodfellas}, \emph{Misery} and \emph{Godfather}. By training with these data points, the model may recommend \emph{Godfather II} to the user. 
If we remove the data point \emph{Godfather} from the training set, indicating that we assume the user has not seen \emph{Godfather} before, and we observe that the recommender, at this time, recommends a different movie to the user, then we can say \emph{Godfather} is the reason of recommending \emph{Godfather II}. To find this counterfactual explanation, we need to accurately know the change to the recommendation -- the item  (\emph{Apt Pupil}) that the model recommends as the replacement of the \emph{Godfather II}.

In this work, we aim at formulating a general framework for counterfactual explainers, regardless of the types of the recommenders, deep neural networks or other machine learning models. Existing counterfactual explainers (e.g., ACCENT and FIA \cite{tran2021ACCENT}) can be considered as specific techniques that belong to this framework with specific choices of the components.
The framework consists of two components: (1) influence score calculation to estimate the influence of the user’s actions on the recommendation results; and (2) a search algorithm to find a minimal set of actions resulting in a different recommendation based on the influence score. 
The most intuitive way to calculate the influence score is to iteratively delete a specific data point (\textit{i.e.,} a user-item interaction pair) and retrain the model. However, the computational cost in this way is intractable. A solution from the existing work~\cite{koh2017understanding,tran2021ACCENT} is the gradient-based method, when the model can be optimized by gradients, where the estimation of the influence score comes from the gradients and the Hessian matrix. 
When the models such as factorization machines (FM) are instead optimized by Bayesian inference or Markov chain Monte Carlo (MCMC) methods,
we introduce a new method with our framework, called data-based method for FM (DB-FM), where we first train the model on the complete data to converge and then continously train it on the training examples without the removed data point. After getting the influence score for each data point, we employ greedy search or iterative greedy search~\cite{tran2021ACCENT} to generate the counterfactual explanation item set. 
Our experiment tasks are to explain two widely-used recommendation models (NCF and FM),
on two popular datasets (Yelp and Movielens). For explanation evaluations, we use two metrics: explanation success percentage and average explanation size following the work~\cite{tran2021ACCENT}.
In experiments, we first investigate the explanation differences among three variants of the counterfactual explainers under the counterfactual explanation framework, which includes ACCENT, FIA, and the data-based method DB-FM. Then we investigate the relationship between the performance of the recommender and the quality of the counterfactual explainer.

We have two preliminary observations. First, the data-based method does not perform comparably well as the gradient-based methods. This is because the gradient-based methods can enforce the models to forget the removed data point, however, the data-based method cannot, though the heavy continuous training no longer uses that point, as the model has seen it before.
So, the problem is still open to create counterfactual explainers for recommender systems whose machine learning models are not optimized by gradients.
Second, the explainers deliver worse performance when the recommendations are more accurate. Having good explanations to correct predictions is harder than having them to wrong predictions. So, the community needs more fine-grained evaluation metrics to measure the quality of counterfactual explanations to recommender systems.

The remaining sections are organized as follows. We discuss the related work to our work in Section 2. In Section 3, we define the problems of counterfactual explanation in recommender systems. A detailed methodology including measuring the influence and generate search results are shown in Section 4. The experiment settings and results are presented in Section 5. We conclude our findings in Section 6.

\section{Related Work}
\label{sec:related}
The interpretation of recommender system predictions is always attracting researchers due to the wide use of modern recommender systems in online websites. Earlier approaches which attempted to interpret the predictions of recommender models tried to impose constraints on them. For example, Seung \textit{et al.} \cite{seung2001algorithms} added a non-negative constraint on matrix factorization models, which aimed to interpret the final preference score as a series of matching score of a certain user to all items. However, such a method cannot generate explicit interpretations. Subsequent approaches adopted neighbor-based interpretations for recommender model predictions. Abdollah \textit{et al.} \cite{abdollahi2016explainable} augmented the objective function with an additional interpretability regularizer on matrix factorization models, and trained the model towards recommending neighboring items to the given user. Heckel \textit{et al.} \cite{heckel2017scalable} built co-clusters of users and items, which helped generate neighbor-based interpretations of the recommended items. Nevertheless, constraint-based interpretation methods may affect the precision of recommendation results. A recent work by Peake \textit{et al.} \cite{peake2018explanation} proposed to perform post-hoc interpretation by learning association rules over the output of a matrix factorization model, but this kind of association rules cannot guarantee to be solvable for any recommendation result.

In contrast to previous interpretability approaches, we mainly follow the line of counterfactual explanation. Counterfactual explanation aims to find counterfactual recommendation results for explaining the relationship between a removed item from past interaction history and the originally recommended item. FIA \cite{cheng2019FIA,tran2021ACCENT} introduced the concept of influence function to estimate the difference after the removal of one history interaction. It leveraged a simple heuristic to find a set of items to remove, making the recommender model to generate a different prediction. ACCENT \cite{tran2021ACCENT} optimized FIA's search algorithm to take the potential counterfactual recommended item into consideration. To get closer to the global minimum, ACCENT is at the expense of reasonable computational time compared to FIA.
In this work, we propose a data-based method for influence score calculation and the search algorithm from ACCENT.
In the broader domain of machine learning, counterfactual explanations are also used in interpreting predictions in computer vision (CV) and natural language processing (NLP). Goyal \textit{et al.} \cite{goyal2019counterfactual} generated counterfactual visual explanations for image classifiaction by replacing a region in the original image. Yang \textit{et al.} \cite{yang2020generating} flipped the unstructured text input to make the model predict a counterfactual class. Critical words to the model prediction in the input sentence are identified and gramatically plausible substitutes were selected as replaced words. Besides the general domain, counterfactual explanation also plays an important role in the financial domain. For example, by generating counterfactual explanations for credit application data, the algorithm is able to tell applicants which factor is critical in the rejection of their credit application, and what could be done to flip the decision \cite{grath2018interpretable}.

\section{Problem Definition}
\label{sec:problem}
Recommender systems are widely used in commercial websites to recommend personalized items (\textit{e.g.}, videos, books, clothing, etc) for a certain user based on a large collection of user-item interaction data (\text{e.g.}, clicking, viewing or ordering). In the era of deep learning, neural recommender systems have become major solutions of item recommendation due to their strong ability of extracting complex features for users and items after sufficient learning. Without loss of generality, we define a neural recommender system as follows:

\begin{dfn}[Neural Recommender Systems]
Suppose we have a set of users $\mathbf{U}$ and a set of items $\mathbf{V}$. We collect a dataset of $n$ user-item interactions $\mathbf{Z} = \{z_1, z_2, ..., z_n\}$, where each data instance $z_i = (u, v)$ represents an recorded interaction between user $u$ and item $v$. A neural recommender system is a neural network $M(\theta)$, where $\theta$ represents the set of learnable parameters in the model $M$. The model is trained on the dataset of user-item interactions $Z$ to predict a preference score $\hat{y}_{u,v}$ for each user-item pair, which represents the estimated user $u$'s interest on item $v$ (rating, click-through probability, etc). Then, the neural recommender model ranks all items in $V$ according to their preference scores given by $M$, and returns a subset of $V$ (usually top-ranked items) as the recommended items to the user.
\end{dfn}

Counterfactual explainability refers to a set of techniques which aim to explore the explainability of machine learning models. They explain a prediction from a machine learning model by calculating a change in the dataset (\textit{e.g.}, removing a data point or modifying some data features) that would cause the underlying machine learning model to make a different prediction. In recommender systems, the counterfactural explanation problem is defined as follows.

\begin{dfn}[Counterfactural Explanation in RecSys]
Suppose there is a set of user-item interaction data $\mathbf{Z} = \{z_1, z_2, ..., z_n\}$, where $z_i = (u, v)$, and a recommender system $M(\theta)$ trained on this dataset. Given a user $u$, the set of items $\mathbf{I}_u$ that had interactions with $u$ and the model's recommendation item $rec$ for the user $u$, the goal of counterfactual explanation is to find a minimal set of items $\mathbf{I}^*_u$, which will lead the model to produce a different recommendation item ${rec}^*$ for the user $u$ if we remove $\mathbf{I}^*_u$ from the dataset $\mathbf{Z}$.
\end{dfn}

In order to achieve this goal, we need to evaluate the influence of removing each item on the model's final recommendation. Hence, the concept of influence function is used for such kind of evaluation.

\begin{dfn}[Influence Function]
Suppose we have a neural recommender model which produces a preference score $\hat{y}_{u,v}$ given any user $u$ and item $v$. For any data object $z = (u, v)$ in the training dataset, the influence function of $z$ on the model's prediction $\hat{y}_{u,v}$ is defined as $I(z, \hat{y}_{u,v})$, where
\begin{equation}
    I(z, \hat{y}_{u,v}) = \hat{y}_{u,v} - \hat{y}^{-z}_{u,v}.
    \label{eq:influence}
\end{equation}
Here, $\hat{y}^{-z}_{u,v}$ refers to the preference score of user-item pair $(u, v)$ after the removal of $z$ from the training dataset.
\end{dfn}

\section{Framework}
\label{sec:method}
In this section, we describe our approaches in generating counterfactual explanation for recommender systems, using the terms and definitions in the previous section.

\subsection{Pipeline Overview}

In general, the pipeline can be broken down to the following steps:

\paragraph{(1) Train the neural recommender model.} The first step is to train a neural recommender model $M(\theta)$ based on the past user-item interactions (training dataset) $\mathbf{Z}$. Here we use one of the popular models, neural collaborative filtering (NCF) \cite{he2017neural} as an example. The inputs to the NCF model are the indices of user $u$ and item $v$. Two embedding tables (one for the users and the other for the items) are used to convert integer indices into dense representation vectors. Then, the user vector and the item vector are sent to a multi-layer neural network, which will generate a preference score $\hat{y}_{u,v}$ for the input user-item pair. 
The model can be optimized through the gradient descent algorithm until convergence.

\paragraph {(2) Produce recommendations for a given user.} After the model is sufficiently trained on the past user-item interactions, given any user $u$, we use the model to calculate preference scores on uninteracted items of the user $u$. Then, we select the top-1 according to preference scores as the recommended item to the user. 

\paragraph {(3) Calculate influence on the recommendation result.} Next, given each data point $z = (u, v)$ in the training dataset, we calculate its influence on the model's recommendation (the recommended item) based on the influence function. The influence of a past user-item interaction $z$ is typically estimated by the difference in model recommendation after removing the data point $z$.

\paragraph {(4) Generate counterfactual explanations.} Based on the influence scores of all past-interacted items of the given user, we try to find a minimal set of items $\mathbf{I}^*_u$, with the removal of which leading to a different recommendation to the given user. The candidates we use to find $\mathbf{I}^*_u$ is from top-$K$ recommended items originally produced by the recommender.

In the following subsections, we will present the details of the framework for the influence score calculation and the search algorithms for the counterfactual explanation generation. We will omit the training and recommendation parts of the recommender system, since they are not the focus of this work. Readers can refer to some popular neural recommender systems for details \cite{youtube2016,he2017neural}.

\subsection{Influence Calculation}
\label{sec:influence}

As is mentioned in Section \ref{sec:problem}, the influence of a data point $z$ on the model's prediction $\hat{y}_{u,v}$ is formulated as $I(z, \hat{y}_{u,v}) = \hat{y}_{u,v} - \hat{y}^{-z}_{u,v}$.
Here, $\hat{y}_{u,v}$ is the model's recommendation based on the original training set of user-item interactions. $\hat{y}^{-z}_{u,v}$ is the model's recommendation based on the modified training set, \text{i.e.}, the dataset after the removal of data point $z$. Ideally, we need to re-train the recommender model after each single item is removed from our training set to get a new set of parameters $\hat{\theta}$, and generate a new preference score $\hat{y}^{-z}_{u,v}$ based on $\hat{\theta}$. However, training a model from scratch multiple times is not efficient. Therefore, we use two methods as estimations: gradient-based estimation and data-based estimation.

\subsubsection{Gradient-Based Estimation}

In gradient-based estimation, we estimate the amount of changes in model's parameters $\theta$ by modifying the weight of a certain training data point. According to \cite{koh2017understanding, tran2021ACCENT}, the influence of upweighting a training data point $z$ by a small amount $\epsilon$ can be estimated as:
\begin{equation}
    \frac{d\theta_{\epsilon,z}}{d\epsilon}\Big|_{\epsilon=0}=-H^{-1}_{\theta}\nabla L(z, \theta),
\label{eq:estimate}
\end{equation}
where $L(z, \theta)$ is the loss function of training data point $z$, and $H^{-1}_{\theta}$ is the Hessian matrix which is computed as $H^{-1}_{\theta}=\frac{1}{n}\sum^{n}_{i=1}\nabla^2_{\theta} L(z, \theta)$.

In our setting where we aim to remove a certain data point $z$, the effect is equivalent to changing its weight in the training dataset from $\frac{1}{n}$ to 0. After setting $\epsilon=-\frac{1}{n}$ in Eq.(\ref{eq:estimate}), we can estimate the new parameters $\hat{\theta}$ as \cite{tran2021ACCENT}:
\begin{equation}
    \hat{\theta} = \theta + \frac{1}{n}H^{-1}_{\theta}\nabla L(z, \theta).
\end{equation}

We can use $\hat{\theta}$ to generate a new recommendation $\hat{y}^{-z}_{u,v}$ for user $u$ because $\hat{\theta}$ is estimated based on the removal of data point $z$.

\subsubsection{Data-Based Estimation}

In the data-based estimation, we aim to obtain new preference score $\hat{y}^{-z}_{u,v}$ by training the model with the training set after the removal of data point $z$. As is mentioned before, it is intractable to train a model from scratch for each possible $z$. Therefore, we propose an alternative solution:
\begin{itemize}
    \item[(1)] Train the model on the entire training data for $T_1$ steps to make the model converge. The set of model parameters at this point is denoted as $\theta$.
    \item[(2)] Remove a certain data point $z$ from the training data.
    \item[(3)] Continuously train the model for additional $T_2$ steps. The set of model parameters $\hat{\theta}$ after this training stage is the estimation.
\end{itemize}

Similar to the gradient-based estimation, we obtain a new recommendation $\hat{y}^{-z}_{u,v}$ for user $u$ based on $\hat{\theta}$.

\subsection{Counterfactual Explanation Generation}

Besides influence score calculation, the other important step is generating the item set for counterfactual generation. Here, our goal is to find the minimal item set $\mathbf{I}^*_u\in \mathbf{I}_u$ to be removed from the training dataset so as to overturn the model's prediction, where $\mathbf{I}_u$ represents the interacted items of user $u$ in the training set. Given the recommended item by the original model as $rec$, we denote the influence score of each data point $z$ on $rec$ as $I(z, \hat{y}_{u,rec})$. According to Eq.(\ref{eq:influence}), $I(z, \hat{y}_{u,rec})=\hat{y}_{u,rec} - \hat{y}^{-z}_{u,rec}$, where $\hat{y}^{-z}_{u,rec}$ is given by the estimation method in Section \ref{sec:influence}. Here, two algorithms are adopted to generate the counterfactual item set $\mathbf{I}^*_u$ based on $I(z, \hat{y}_{u,rec})$.

\subsubsection{Greedy Search}

In greedy search for counterfactual explanation, we first calculate $I(z, \hat{y}_{u,rec})$ for each data point $z\in \mathbf{I}_u$ in the training set. Then we greedily search for the data points with the largest $I(z, \hat{y}_{u,rec})$ score and add the corresponding $z$ to the item set $\mathbf{I}^*_u$. It continues until the originally recommended item $rec$ is no longer the top-1 recommendation by the estimated re-trained model $M(\hat{\theta})$, \textit{i.e.}, the recommended item $rec$ is replaced by another item $rec^{*}$.

\subsubsection{Iterative Greedy Search}

In greedy search, we do not consider all items in the top-$K$ recommended items as potential $rec^{*}$ to replace the original $rec$. It may lead to suboptimal solutions in practice.
Therefore, in iterative greedy search, we iteratively consider each possible candidate item as a potential $rec^{*}$, and conduct greedy search on each potential $rec^{*}$ respectively. A globally minimal item set $\mathbf{I}^*_u$ will be obtained after all iterations.

First, we first derive the influence of removing a data point $z$ on the score difference between two items $v$ and $w$ as:
\begin{equation}
    \begin{aligned}
    I(z,&\hat{y}_{u,v}-\hat{y}_{u,w})=(\hat{y}_{u,v}-\hat{y}_{u,w})-(\hat{y}^{-z}_{u,v}-\hat{y}^{-z}_{u,w}) \\
    &=(\hat{y}_{u,v}-\hat{y}^{-z}_{u,v})-(\hat{y}_{u,w}-\hat{y}^{-z}_{u,w})=I(z,\hat{y}_{u,v})-I(z,\hat{y}_{u,w}).
    \end{aligned}
\end{equation}

\begin{table*}[]
\caption{Component settings of different models in our experiments.}
\label{tab:models}
\begin{tabular}{llll}
\toprule
\textbf{Model}  & \textbf{Influence Score Calculation} & \textbf{Search Algorithm}        & \textbf{Base Recommender Model}               \\ 
\midrule
ACCENT \cite{tran2021ACCENT} & Gradient-based              & Iterative Greedy Search & Neural Collaborative Filtering (NCF) \\
FIA \cite{cheng2019FIA}    & Gradient-based              & Greedy Search           & Neural Collaborative Filtering (NCF) \\
DB-FM  & Data-based            & Iterative Greedy Search & Factorization Machine (FM)           \\ 
\bottomrule
\end{tabular}
\end{table*}

Next, assume that we have a set of candidates $I_{cand}$ from the originally recommended top-$K$ items.
For each item $i \in I_{cand}$, there is a difference between the preference scores of the actual recommended item $rec$ and item $i$ from the original model:

\begin{equation}
    \textit{diff}=\hat{y}_{u,rec}-\hat{y}_{u,i}>0.
\end{equation}

Then we sort all the data points $z\in \mathbf{I}_u$ according to their influence on items $rec$ and $i$, \textit{i.e.}, $I(z,\hat{y}_{u,rec}-\hat{y}_{u,i})$. We greedily select a data point $z$ starting from the one with largest influence, and apply the influence score to update the estimated difference after the removal of $z$:
\begin{equation}
    \textit{diff}=\textit{diff}-I(z,\hat{y}_{u,rec}-\hat{y}_{u,i}).
\end{equation}

We continue until reaching a point where $\textit{diff}<0$. It means the removal of data points $\mathbf{I}^*_u$ has overturned the model's recommendation. Finally, we select the item $i$ with minimal size of counterfactual item set $\mathbf{I}^*_u$ as $rec$. Therefore, the recommendation $rec^{*}$ can be counterfactually explained by $rec^{*}$ and $\mathbf{I}^*_u$.

\section{Experiments}
\label{sec:experiments}
In the experiments, we first define the explanation metrics following the work in~\cite{tran2021ACCENT} to measure the performance of different explainers. Then we implement multiple counterfactual explanation methods, including ACCENT~\cite{tran2021ACCENT}, FIA~\cite{tran2021ACCENT,cheng2019FIA} and a data-based method DB-FM. Besides, we further analyze how the model's recommendation performance (\textit{i.e.}, how well the model learned from the dataset) affects the quality of counterfactual explanation.

\begin{table*}
\centering
\caption{Results on two datasets: {MovieLens} and {Yelp}. ACCENT and FIA share the same recomendation results (MSE) because their base recommender models are both NCF. We consider different numbers of $K$ for top-$K$ recommendations. ESP: Explanation Success Percentage (\%) ($\uparrow$: bigger is better). AES: Average Explanation Size ($\downarrow$): smaller is better).}
\label{tab: explanation results}
\textbf{MovieLens dataset:} \\
\begin{tabular}{lc|rr|rr|rr}
\toprule
{\multirow{2}*{Explainer}} & {\multirow{2}*{MSE ($\downarrow$)}} & \multicolumn{2}{c|}{{$K=5$: Top-$5$ recommendation}} & \multicolumn{2}{c|}{{$K=10$: Top-$10$ recommendation}} & \multicolumn{2}{c}{{$K=20$: Top-$20$ recommendation}} \\ \cline{3-8}
& & \multicolumn{1}{c}{ESP ($\uparrow$)} & \multicolumn{1}{c|}{AES ($\downarrow$)} & \multicolumn{1}{c}{ESP ($\uparrow$)} & \multicolumn{1}{c|}{AES ($\downarrow$)} & \multicolumn{1}{c}{ESP ($\uparrow$)} & \multicolumn{1}{c}{AES ($\downarrow$)} \\
\midrule
ACCENT \cite{tran2021ACCENT} & \textbf{0.022} & \textbf{54} & 9.759 & \textbf{54} & 9.759 & \textbf{53} & 9.906 \\
FIA \cite{cheng2019FIA} & \textbf{0.022} & 50 & 11.900 & 50 & 11.900 & 49 & 12.061 \\
DB-FM & 0.375 & 10 & \textbf{1.200} & 9 & \textbf{1.222} & 7 & \textbf{1.286} \\
\bottomrule
\end{tabular} \\
\vspace{0.1in}
\textbf{Yelp dataset:} \\
\begin{tabular}{lc|rr|rr|rr}
\toprule
{\multirow{2}*{Explainer}} & {\multirow{2}*{MSE ($\downarrow$)}} & \multicolumn{2}{c|}{{$K=5$: Top-$5$ recommendation}} & \multicolumn{2}{c|}{{$K=10$: Top-$10$ recommendation}} & \multicolumn{2}{c}{{$K=20$: Top-$20$ recommendation}} \\ \cline{3-8}
& & \multicolumn{1}{c}{ESP ($\uparrow$)} & \multicolumn{1}{c|}{AES ($\downarrow$)} & \multicolumn{1}{c}{ESP ($\uparrow$)} & \multicolumn{1}{c|}{AES ($\downarrow$)} & \multicolumn{1}{c}{ESP ($\uparrow$)} & \multicolumn{1}{c}{AES ($\downarrow$)} \\
\midrule
ACCENT \cite{tran2021ACCENT} & \textbf{0.113} & \textbf{27} & 11.667 & \textbf{29} & 11.862 & \textbf{27} & 12.667 \\
FIA \cite{cheng2019FIA} & \textbf{0.113} & 22 & 10.591 & 22 & 10.136 & 20 & 11.05 \\
DB-FM & 0.375 & 9 & \textbf{1.000} & 7 & \textbf{1.000} & 5 & \textbf{1.000} \\
\bottomrule
\end{tabular}
\end{table*}


\subsection{Experiment Settings}

\paragraph{Datasets}
We use two popular recommendation datasets -- Yelp~\cite{asghar2016yelp} and MovieLens\footnote{\url{https://grouplens.org/datasets/movielens/}} for the experiments:
\begin{itemize}
\item \textit{MovieLens dataset}: We use MovieLens 100K dataset which includes 100K ratings on a 1-5 scale for 1682 movies given by 943 users. We remove all users with less than 10 actions in the dataset to filter out possibly noisy data. The filtered dataset contains 453 users and 1654 movies. There are a total of 61,054 user-item interactions. The density of this user-item graph is 8.44\%\footnote{Density is number of ratings divided by number of users and by number of items.}.
\item \textit{Yelp dataset}: The Yelp dataset contains ratings on a 1-5 scale from 2M users on 160k restaurants. The dataset also contains profiling features for users and restaurants. User features include his/her ID, nickname, received votes, average stars, etc. Item features consist of its ID, name, address, city, star rating and category. We first filter the dataset to select 1200 most active users and 1200 most popular restaurants from all the reviews. Then we use their intersections to construct our dataset. There are a total of 10,827 user-item interactions in this Yelp subset. The density of the user-item graph in this subset is 6.31\%.
\end{itemize}

\paragraph{Models}

In our experiments, we implement two kinds of counterfactual explanation methods, ACCENT~\cite{tran2021ACCENT}, FIA~\cite{tran2021ACCENT, cheng2019FIA} and DB-FM, based on our framework described in Section 4. ACCENT and FIA are based on different combinations of influence score estimation methods and search algorithms. They both leverage gradient-based influence score estimation but vary in search algorithms. DB-FM is based on data-based influence score estimation and the interactive greedy search algorithm. We adapt ACCENT and FIA on neural collaborative filtering (NCF) as the base recommender model, while DB-FM is based on factorization machine (FM). A detailed list of model component settings is listed in Table~\ref{tab:models}.

During training, NCF-based methods (ACCENT and FIA) are optimized by gradient descent, while DB-FM is optimized with Markov Chain Monto Carlo search. We use mean squared error (MSE) to evaluate the performance of recommender models.

\paragraph{Evaluation Metrics}

For a particular user, the counterfactual explainer outputs the minimal item set to remove $\mathbf{I}^*_u$ for counterfactual explanation, as well as the replaced recommendation item $rec^*$ from the original top-$K$ recommendations. We re-train the model without $\mathbf{I}^*_u$ in the training set to see if the removal actually takes effect to replace the original top-1 recommendation with the item $rec^*$. We use two explanation metrics to measure the performance of different counterfactual explainers following the work \cite{tran2021ACCENT}:

\begin{itemize}
\item \textit{Explanation Success Percentage (ESP)}: An explanation is counted as a successful explanation if the top-1 recommendation $rec$ indeed change to the predicted $rec^*$ after the model is re-trained on the training set without $\mathbf{I}^*_u$. Bigger ESP means better explanations.
\item \textit{Average Explanation Size (AES)}: The average size of $\mathbf{I}^*_u$ over all the users whose recommendations are successfully explained. Smaller ESP means better explanations.
\end{itemize}

In experiments, the choice of $K$ may affect the scores of these metrics. The value of $K$ controls how many candidate items to be considered as potential $rec^*$ to replace the original recommendation. We report results while choosing different values of $K$ in Table~\ref{tab: explanation results}.

\subsection{Results of Counterfactual Explanations}
In this section, we report the results for our experiments.

\subsubsection{Recommendation Performance}

We first report the recommendation performance of these models. This is mainly related to how well the base recommender model is trained on these datasets. The results are shown in Table~\ref{tab: explanation results}. Generally, the NCF model learns better than the FM model, with lower MSE on both datasets. To generate the test set for counterfactual explanation methods, we first sample 100 users with their top-1 recommendations from the base recommender  models, then run the counterfactual explanation methods to get $rec^*$ from the top-5, top-10, or top-20 candidate recommendations as well as find the minimal set of $\mathbf{I}^*_u$.

\subsubsection{Explanation performance}

Next, we summarize the counterfactual explanation results in the Table~\ref{tab: explanation results}. Summarizing scores from two datasets and three different values of $K$, we observe that ACCENT generally performs better than FIA and DB-FM in explanation success percentages.Compared to the FIA, which has relatively larger explanation set sizes, the explanation from the ACCENT has a smaller size and is closer to the most vital reason for recommendation results. Besides, we observe that the DB-FM model usually takes aggressive counterfactual explanations using less explanation items, which may result in poor explanation success percentages.

The data-based method does not perform comparably well as the gradient-based methods. This is because the gradient-based methods can enforce the models to forget the removed data point, however, the data-based method cannot, though the heavy continuous training no longer uses that point, as the model has seen it before.
So, the problem is still open to create counterfactual explainers for recommender systems whose machine learning models are not optimized by gradients.

\subsubsection{Discussion on Counterfactual Explanation Strategies}

Comparing ACCENT to FIA, we find that the more complicated iterative greedy search algorithm indeed performs better than the simpler greedy search. By iteratively considering which item to replace the original recommendation, the algorithm is more likely to reach the global minimal explanation set instead of sliding to suboptimal solutions. If we ignore the impact of base recommender models, we find that the influence score estimation method impacts both the explanation success percentage and the explanation size. Data-based estimation may result in more aggressive explanations.

\subsubsection{Discussion on the Shortcomings of Iterative Greedy Search.}
It is obvious that iterative greedy search performs better than greedy search if we compare \accentt to \fia. As stated in Section~\ref{sec:method}, iterative greedy search finds the replaced item $rec^*$ by calculating the explanation size $\mathbf{I}^*_u$ for each potential replacement and get the global minimum. However, we find from the experiments that smaller explanation sizes may also indicate more risky explanations, resulting in worse explanation success percentages.

\subsection{Analysis on Embedding Size}

\begin{figure}[t]
    \centering
    {\includegraphics[width=\linewidth]{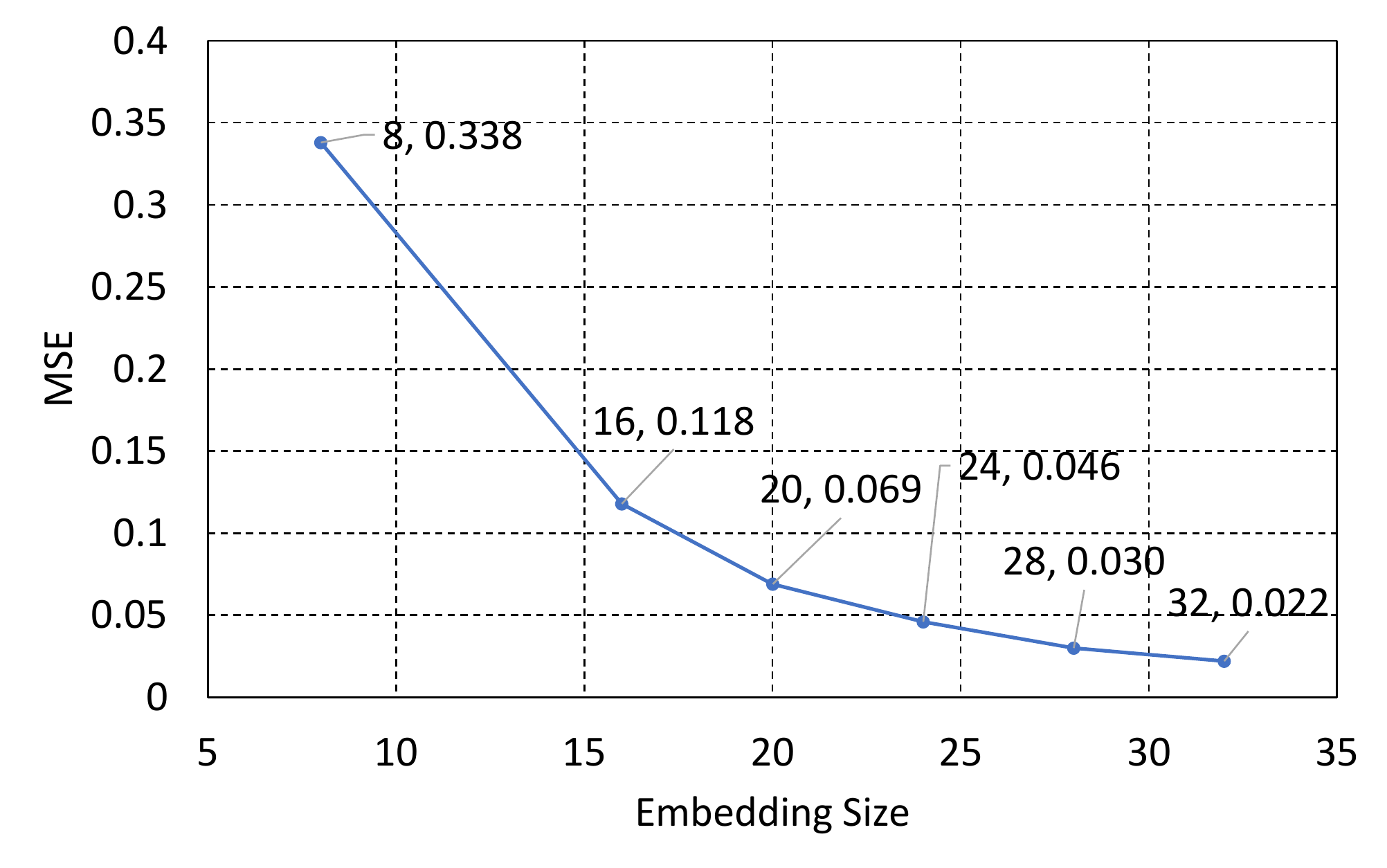}\label{fig:mse_emb}}
    \caption{Training MSE rate of the NCF model under different embedding dimensions: $d \in \{8, 16, 20, 24, 28, 32\}$.}
    \label{fig: mse_embd}
\end{figure}

Diving deeper into experimental results, we are interested to how the performance of the base recommender model affects the quality of counterfactual explanations. That is, if a model generates perfect recommendations, an optimal explanation method should recover the causal relationship discovered by the recommender model and perform best on all explanation metrics. However, if the model is not well trained, we are curious about whether it impacts the quality of the explanation methods. In real applications, both the intrinsic design (such as hyperparameters) and the data distribution may affect the performance of models, and indirectly affect the performance of the explanation method as well. Therefore, we modify some of the settings in the base recommender model to obtain model checkpoints with different recommendation performances, and again test the ACCENT explanation method on these model checkpoints. To be specific, we test NCF models with different embedding dimensions, one of the major factors that may affect the performance of a neural recommender model. Here, we use MSE to measure the performance of recommender models.

We show the MSE scores of the base NCF model under different hyper-parameter setting in Figure \ref{fig: mse_embd}.
Specifically, we first vary the embedding dimensions of NCF in the range of \{8, 16, 20, 24, 28, 32\}.
Clearly higher embedding dimensions lead to more complex models, which have lower MSE rates after sufficient training.
\begin{figure}[t]
    \centering
    {\includegraphics[width=\linewidth]{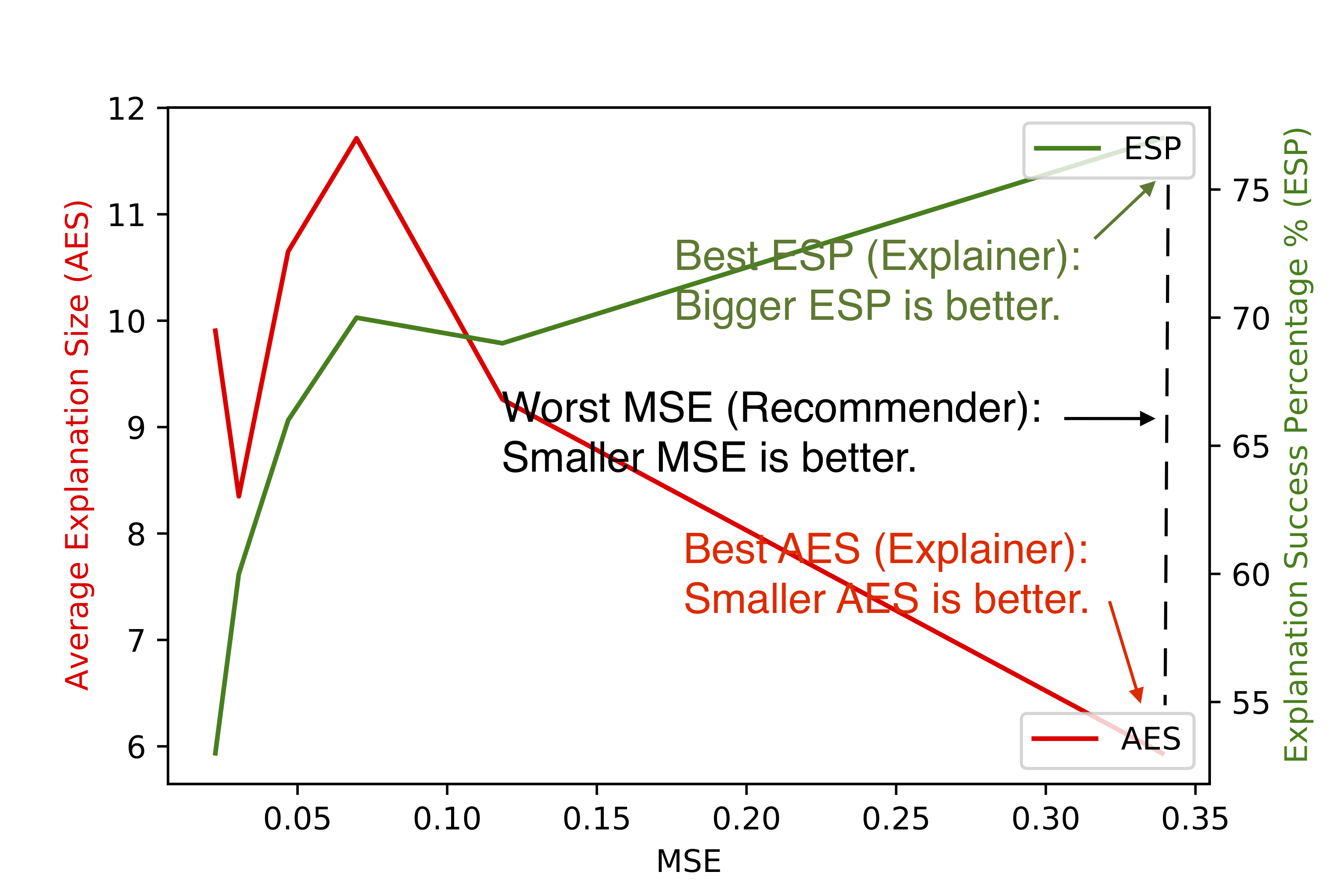}\label{fig:mvlen_accent_emb}}
    \caption{How explanation metrics changes on the MovieLens dataset using NCF-based \accentt by varying embedding dimension $d \in \{8, 16, 20, 24, 28, 32\}$.}
    \label{fig:mvlen accent emb and lr}
\end{figure}

\subsection{Analysis on Explainer Performance vs. Recommender Performance}

According to Figure \ref{fig:mvlen accent emb and lr}, we observe that the quality of counterfactual explanations is indeed affected by the performance of the base recommender model. Note that smaller explanation sizes and higher success percentages indicate better explanations. When the model is generally well trained (with MSE smaller than 0.1), generating counterfactual explanations tends to be more difficult than the model is less well trained (higher MSE). If the model performs relatively badly (with MSE larger than 0.1), generating counterfactual explanations will instead become easier. One possible explanation for these phenomena is that when the model performs really well, gradient-based estimations of influence scores used by ACCENT need to be more accurate to get good explanations, which is hard. When the model performs really badly, the confidence of the original recommendation result tends to be pretty low, then the ACCENT method is able to overturn the model's recommendation easily.
The explainers deliver worse performance when the recommendations are more accurate. Having good explanations to correct predictions is harder than having them to wrong predictions. So, the community needs more fine-grained evaluation metrics to measure the quality of counterfactual explanations to recommender systems.

\section{Conclusions}
\label{sec:conclusion}
In this work, we explored multiple settings of counterfactual explanations in recommender systems. Based on experiments on two benchmark datasets, we found that the ACCENT method, equipped with gradient-based influence score estimation and iterative greedy search, achieved competitive results when applied on the NCF model. However, the iterative greedy search algorithm might also lead to risky explanations with smaller explanation set size but potentially higher error rates. We also investigated how the performance of the base recommender model affects the quality of counterfactual explanations.

\begin{acks}
This work is supported in part by ONR N00014-22-1-2507 and Notre Dame International Research Grant.
\end{acks}

\balance
\bibliographystyle{ACM-Reference-Format}
\bibliography{ref}


\end{document}